\documentclass{ws-ijmpd}

\def\ApJ{{\it Astrophys. J.} }

\def\ApP{{\it Astropart. Phys.} }
\def\AA{{\it Astron. \& Astroph.} }

\def\MNRAS{{\it Month. Not. Roy. Astr. Soc.} }
\def\Nature{{\it Nature} }
\def\NewAR{{\it New Astron. Rev.} }

\def\PRL{{\it Phys. Rev. Letters} }

\def\Science{{\it Science} }

\begin{document}

\markboth{P.L. Biermann, J.K. Becker, L. Caramete, L.\'A. Gergely, I.C. 
Mari\c{s}, A. Meli, V. de Souza, T. Stanev}
{Ultra high energy cosmic rays}

%
%

\title{ACTIVE GALACTIC NUCLEI WITH STARBURSTS: SOURCES FOR ULTRA HIGH ENERGY
COSMIC RAYS}

\author{P.L. Biermann$^1$\footnote{Also at Univ. of Bonn, Univ. of Alabama,
Tuscaloosa, AL, Univ. of Alabama at Huntsville, and at KIT Karlsruhe},
J.K.~Becker$^2$\footnote{also at the University of Dortmund, and
from June 2009 also at the University of Bochum},
L.~Caramete$^3$, L.\'A.~Gergely$^4$, I.C.~Mari\c{s}$^5$,
A.~Meli$^6$, V.~de~Souza$^7$, T.~Stanev$^8$ }
\address{$^1$MPI for Radioastronomy, D-53121 Bonn, Germany
 (www.mpifr-bonn.mpg.de/div/theory), $^2$Institution f{\"o}r
 Fysik, G{\"o}teborgs Univ., Sweden, $^3$Institute for Space Sciences, 
Bucharest, Romania, $^4$Phys. Dept., Univ. of Szeged,
 Szeged, Hungary, $^5$Inst. Nucl. Phys. FZ, Karlsruhe Inst. of Techn.
 (KIT), Germany, $^6$Erlangen Center for Astroparticle Physics,
Univ. Erlangen-N{\"u}rnberg, Germany,
 $^7$Universidade de S$\tilde{a}$o Paulo, Instituto de F\'{\i}sica de
S$\tilde{a}$o Carlos, Brazil, $^8$Bartol Research Inst., Univ. of 
Delaware, Newark, DE, USA}
 










%

\maketitle


\begin{abstract}
Ultra high energy cosmic ray events presently show a spectrum, which we 
interpret here as galactic cosmic rays due to a starburst in the radio 
galaxy Cen A pushed up in energy by the shock of a relativistic jet.  
The knee feature and the particles with energy immediately higher in 
galactic cosmic rays then turn into the bulk of ultra high energy cosmic 
rays.  This entails that all ultra high energy cosmic rays are heavy 
nuclei.  This picture is viable if the majority of the observed ultra 
high energy events come from the radio galaxy Cen A, and are scattered 
by intergalactic magnetic fields across most of the sky.
\end{abstract}


\section{Introduction}    

The origin of ultra high energy cosmic ray has been a riddle since their 
discovery 1963 by Linsley (Ref.~\refcite{Linsley63}).  After many 
experiments and measurements we are beginning to accumulate reasonable 
statistics with the Auger array, and have a very good spectrum, shown in 
our figure below (see Ref.~\refcite{Augerspectrum}).

As there is a weak correlation with active galactic nuclei, various 
classes of such objects have come under consideration again.  Radio 
galaxies have been argued to be sources for some time (see 
Ref.~\refcite{GS63SAJ}), but none in our cosmic neighborhood readily 
accelerates protons to such high energy, as derived from the jet 
energetics (see Ref.~\refcite{Lovelace76}).  Heavy nuclei seem to be 
required.  Here we outline such a scenario (see Ref.~\refcite{Malfa08}), 
and give a spectral fit in the figure below.

\section{Starburst and relativistic jet}

When two galaxies merge  (see Ref.~\refcite{spinflip}), we first have a 
starburst, and second a central activity that outshines the young stars, 
and also disperses much of the interstellar material.  Usually such a 
merger is also accompanied by a spinflip of the more massive black hole, 
due to the dominant relative orbital spin, when two black holes merge.  
This implies that a new jet forms and bores a new channel through the 
environment, freshly filled with galactic cosmic rays. These new 
galactic cosmic rays have a relatively flat spectrum to the bend, 
commonly referred to as the knee, where the spectrum turns down.  
Gallant \& Achterberg (Ref.~\refcite{GA99}) have shown that the shock in 
a single interaction pushes the particle energies up by $\Gamma_j^{2}$, 
where $\Gamma_j$ is the the Lorentz factor  of the shock.  Studies of 
radio interferometry and other work has suggested that in the spine of 
the jet this Lorentz factor can reach 50.  Therefore the spectral 
distribution near the knee (see Ref.~\refcite{CRIV}) may be pushed up in 
energy by a factor up to 2500.  This then clearly provides ultra high 
energy particles.  The only source in our cosmic neighborhood capable of 
doing this is the radio galaxy Cen A.


Obviously, we do need scattering by magnetic fields (see 
Ref.~\refcite{RKB98,RKCD08,Dasetal08}) to explain a spectrum built from 
events all over the sky, but coming from a single source.  In fact, 
since their MHD cosmological simulations already give some large angle 
scattering for protons, and we here argue for heavy nuclei, those models 
would produce an almost perfectly isotropic sky.  However, as shown by 
inconsistencies in the magnetic field determinations in our Galaxy (see 
Ref.~\refcite{Beck03}), the medium is probably highly filamentary, 
strongly reducing the angular scattering.  This is reasonable for highly 
super-Alfv{\'e}nic turbulence injected by super-nova explosions.  In the 
intergalactic medium super-Alfv{\'e}nic turbulence may be due to radio 
galaxies.  Thus the scattering may be moderate even for heavy nuclei, 
explaining the large scale nearly isotropic sky distribution, but also 
allowing many events to come from directions close to Cen A.

\section{The spectrum}

We have taken the spectrum for galactic cosmic rays 
(Ref.~\refcite{CRIV}), pushed it up to match the turnoff feature near 40 
EeV and fitted it to the Auger data, as published 
(Ref.~\refcite{Augerspectrum}).  We match the slope below the Fe-edge at 
40 EeV, the step down, and the slope beyond 40 EeV.  Obviously it is 
possible to improve the match, and so we tested modifying the 
abundances; we note that the abundances of the various chemical elements 
in the winds of Wolf-Rayet stars and Red Super Giant stars depend on the 
initial abundances, but most of all on the zero age masses of the stars, 
and so on their mass distribution:  Therefore it cannot a priori be 
expected, that the abundances in cosmic rays in our Galaxy match those 
in another galaxy. A decrease by 20 percent the abundances of the two 
element groups Ne-S and Cl-Mn improves the match below 40 EeV to near 
perfect.  Due to the small distance of the radio galaxy Cen A, we do not 
expect any strong effect from propagation; however, since some particles 
are required in our model to make detours, that clearly is another 
simplification.  We show the match without taking any of this into 
account (Ref.~\refcite{Augerspectrum,CRIV}):

\begin{figure}[h!]
\centering
\includegraphics[scale=0.65]{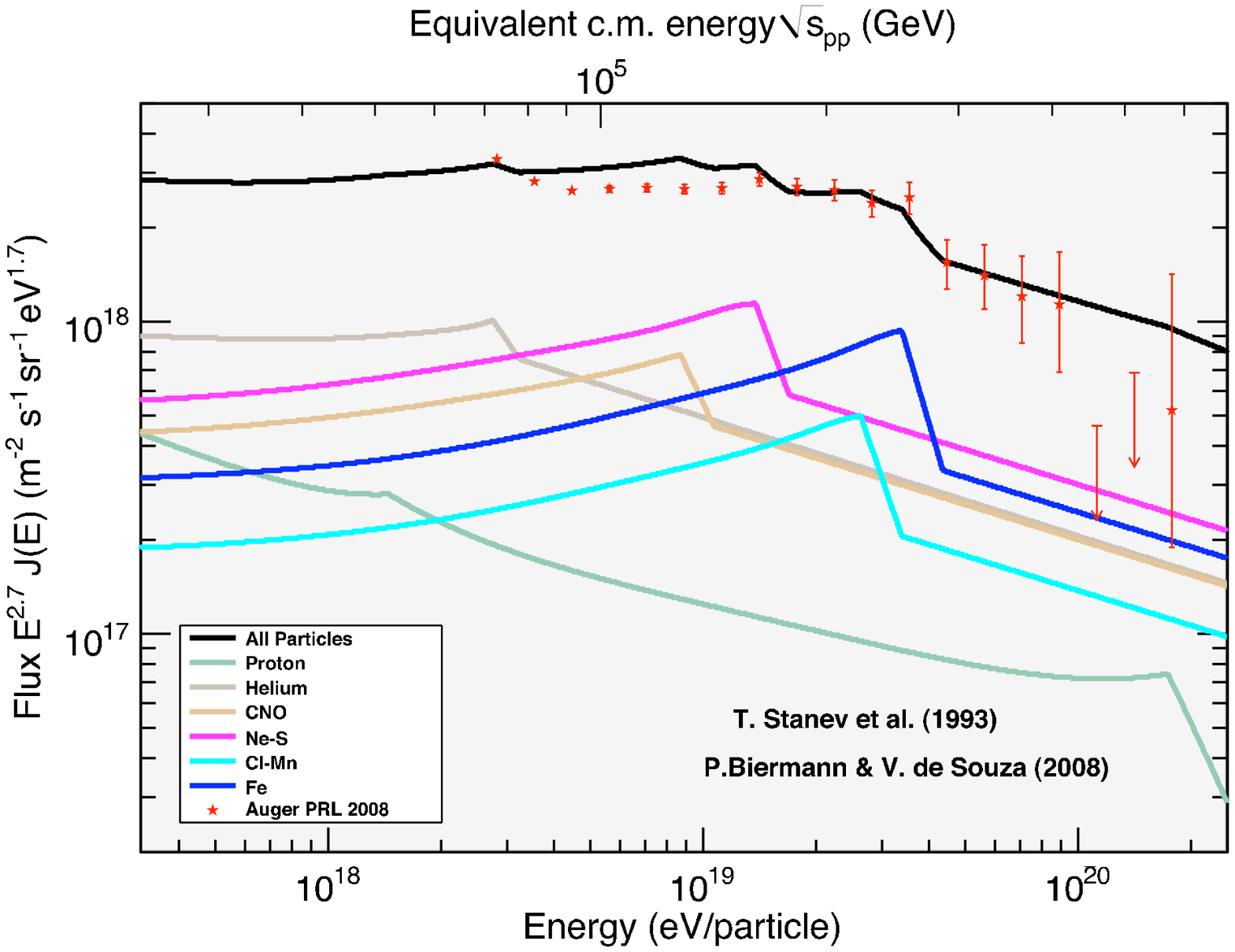}
\vspace*{8pt}
\caption{Testing the spectrum in paper CR-IV with Auger data, with all 
six element groups unmodified.  }
\end{figure}
\label{Spectrum}


Taken literally, this interpretation of the data suggests that beyond 40 
EeV the two dominant element groups are Ne - S and Iron, all coming from 
Cen A.


\section{Discussion, conclusions and predictions}

As interactions of photons with heavy nuclei give very much lower fluxes 
of high energy photons and neutrinos, we wonder, whether this argument 
would not reduce the high energy neutrino fluxes to below any possible 
detection limit (Ref.~\refcite{BB09}).  On the other hand, we note that 
we expect high energy neutrinos from flat spectrum radio sources, which 
have their current relativistic jet pointed at the Observer on Earth.  
All such sources are highly variable, and so also the interactions may 
run right into saturation for many flares, producing very high neutrino 
fluxes, but for a very short time.

If there are any protons among the ultra high energy particles, they 
should point better towards their sources; very few events point to the 
radio galaxy Cen A (note, that Cen A extends 10 degrees on the sky: 
Ref.~\refcite{JunkesCenA}).  Those that do appear to come from inside 
the radio source extent cannot be protons and also be attributed to Cen 
A, since Cen A cannot give so high energies to protons, by the Poynting 
flux limit (see Ref.~\refcite{Lovelace76}) and using its jet power 
(Ref.~\refcite{Antonucci03}; this is also true, if protons initially 
accelerated get converted to neutrons that decay back to protons 
(Ref.~\refcite{Rachen08}).

This model predicts:  a) ultra high energy cosmic rays are mostly the 
elements Ne - S and Iron; b)  most of events come from the radio galaxy 
Cen A; and c) we do need highly filamentary magnetic structures in the 
intergalactic medium which scatter to large angles without completely 
isotropizing the paths of the events.
Further statistics should confirm spectrum, scattering distribution, and 
chemical composition.

\section*{Acknowledgments}

Support for work with PLB has come from the AUGER membership and theory 
grant 05 CU 5PD 1/2 via DESY/BMBF and VIHKOS.  Support for JKB comes 
from the DFG grant BE-3714/3-1 and from the IceCube grant BMBF (05 
A08PE1).  V. de Souza is supported by FAPESP (2008/04259-0). Support for 
TS comes from DOE grand UD-FG02-91ER40626.




\begin{thebibliography}{99}    
\bibitem{Linsley63} Linsley, J., \PRL  {\bf 10}, 146 - 148 (1963)

\bibitem{Augerspectrum}  Auger-Coll., \PRL {\bf 101}, ms. 061101 (2008)

\bibitem{GS63SAJ}  V. L. Ginzburg \& S. I. Syrovatskii, {\it Astron. 
Zh.}  {\bf 40} (1963) 466 - 476 ; transl. in {\it Sov. Astron. A.J.}  {\bf 7} (1963) 357 - 364

\bibitem{Lovelace76}  R.V.E. Lovelace,  \Nature {\bf 262} (1976) 649

\bibitem{Malfa08}  P. L. Biermann et al., in  proc. CRIS2008, Malfa, 
Salina Island, Italy, Ed. A. Insolia, (2008) in press; arXiv: 0811.1848v3

\bibitem{spinflip}  L. Gergely \& P.L. Biermann,  in press \ApJ (2009); arXiv:0805.4582

\bibitem{GA99}   Y.-A. Gallant, \&  A. Achterberg, \MNRAS {\bf 305} (1999) L6 - L10

\bibitem{CRIV} T. Stanev, P.L. Biermann \& T.K. Gaisser, \AA {\bf 274} (1993) 902 ; astro-ph/9303006 (CR-IV)

\bibitem{RKB98}  D. Ryu, H. Kang \& P.L. Biermann, \AA  {\bf 335} (1998) 19 - 25; astro-ph/9803275

\bibitem{RKCD08}  D. Ryu et al., 2008 \Science  {\bf 320} (2008) 909
    
\bibitem{Dasetal08}  S. Das et al., \ApJ {\bf 682} (2008) 29 - 38
    
\bibitem{Beck03} R. Beck et al. \AA {\bf 411} (2003) 99 - 107

\bibitem{BB09}  J. K. Becker \& P. L. Biermann, \ApP {\bf 31} (2009) 138; arXiv:0805.1498

\bibitem{JunkesCenA}  Junkes, N., et al. \AA {\bf 269} (1993) 29 - 38; erratum:  \AA {\bf 274} (1993) 1009

\bibitem{Antonucci03}  Whysong, D., Antonucci, R., \NewAR  {\bf 47} (2003) 219 - 223

\bibitem{Rachen08}  Rachen, J.P., XXth Rencontres de Blois, "Challenges in Particle Astrophysics", arXiv:0808.0349 (2008)

\end{thebibliography}
\end{document}